\documentclass[12pt,epsf,psfig,aps,eps]{article}
\usepackage{fullpage}
\usepackage{graphicx}
\input epsf

\begin{document}

\Large
\centerline{\bf Multiscale Algorithms for Eigenvalue Problems}
\normalsize
\vspace{.5cm}
\centerline{Nimal Wijesekera$^1$, Guogang Feng$^2$, Thomas L. Beck$^{2*}$}
\centerline{\it $^1$Department of Physics}
\centerline{\it $^2$Department of Chemistry}
\centerline{\it University of Cincinnati}
\centerline{\it Cincinnati, OH 45221-0172}
\centerline{thomas.beck@uc.edu}
\centerline{*corresponding author}
\centerline{\today}

\vspace{1.cm}
\large
\centerline{\bf Abstract}
\normalsize
\vspace{.5cm}

Iterative multiscale methods for electronic structure calculations
offer several advantages for large-scale problems. Here we
examine a nonlinear full approximation scheme (FAS) multigrid
method for solving fixed potential and self-consistent
eigenvalue problems. In principle, the expensive 
orthogonalization and Ritz projection 
operations can be moved to coarse levels, thus substantially
reducing the overall computational expense. Results of 
the nonlinear multiscale approach are presented for simple
fixed potential problems and for self-consistent 
pseudopotential calculations on large molecules. It is shown
that, while excellent efficiencies can be obtained for 
problems with small numbers of states or well-defined eigenvalue
cluster structure, the algorithm in its original form
stalls for large-molecule problems with tens of occupied levels. 
Work is in progress to attempt 
to alleviate
those difficulties.  

\vspace{1.cm}
\large
\centerline{\bf Introduction}
\normalsize
\vspace{.5cm}

Electronic structure methods for large-scale problems can 
be divided into three general categories:  plane-wave \cite{mpay92},
traditional basis set \cite{mcha00,dsan97}, 
and real-space methods \cite{tbec00}.  
Real-space methods result in a banded Hamiltonian, 
through finite difference 
\cite{jche94,ebri96,jber97,kiye95,jwan00,tbec01,mhei01,fanc99,thos97,ilee00,nmod97,jfat99,jfat00}, 
finite element \cite{etsu98,cgor97,jpas99}, 
or wavelet \cite{tari99} representations of the Laplacian operator. Gaussian
basis sets lead to a smaller total matrix size of the Hamiltonian
(relative to real-space methods), 
but the matrix is less banded. The plane-wave basis set,
on the other hand, is completely delocalized in real space. 
The bandedness of the Hamiltonian in real-space methods is 
advantageous in several respects.  First, it leads to ease of 
developing parallel codes. Second, methods developed recently
which scale linearly with the system size generally rely on 
localization of the orbitals and real-space 
methods `mesh' well with those algorithms \cite{sgoe99}.  
Third, efficient
multiscale methods accelerate convergence by decimating errors
over a wide range of length scales. Fourth, finite clusters
or periodic systems can be treated with equal effort. Finally,
local mesh refinements can be incorporated without degrading
the efficiency of the solver \cite{jfat99,tbec99}. 
Several groups have developed real-space solvers for the 
Kohn-Sham equations of density functional theory (DFT) in 
the last decade \cite{tbec00}. Multigrid (MG) methods have been employed 
extensively to accelerate the convergence rate 
\cite{ebri96,jwan00,mhei01,fanc99,thos97,ilee00,nmod97,jfat99,jfat00}.
 
A central feature of the Kohn-Sham problem is its nonlinearity.
It is nonlinear in two
respects. First, the eigenvalue problem itself is nonlinear
since one solves for both the eigenvalues and eigenfunctions.
Second, the self-consistent potential depends nonlinearly on 
the charge density obtained from the squares of the 
eigenstates.  Therefore, it can be expected that 
nonlinear multigrid methods will lead to increased efficiencies.
This has indeed been observed previously in studies which compare 
a full approximation scheme (FAS) nonlinear approach to linearized 
MG methods \cite{jwan00,fanc99}. 

The most costly operation in the nonlinear 
eigenvalue approach of Brandt {\it et al.} \cite{jwan00,abra83} is
the Ritz projection (preceeded by Gram-Schmidt orthogonalization) on 
the fine level. If $q$ orbitals span the whole physical domain, with
$N_g^h$ grid points on the fine level (labelled by $h$), 
this step of the algorithm scales as $q^2N_g^h$.
In the present paper, we carry the nonlinear
FAS scheme further by implementing an
extension of the Brandt {\it et al.} algorithm proposed by 
Costiner and Ta'asan \cite{scos95a,scos95b}.  
In their approach, the orthogonalization
and Ritz projection are moved to coarse levels within the FAS
strategy. This results in an 8-fold decrease in computational
cost for the Gram-Schmidt/Ritz operation
for each coarser level in three dimensions.
We discuss relevant details of their method, and then 
present numerical results on fixed potential and self-consistent
eigenvalue problems related to atomic and molecular structure.

\vspace{1.cm}
\large
\centerline{\bf Algorithms}
\normalsize
\vspace{.5cm}

Multigrid methods accelerate the convergence rate of iterative
relaxation solvers for partial differential equations by decimating
errors on a wide range of length scales \cite{abra77}; 
it is the long wavelength
modes of the error which degrade single-level relaxation efficiency.
Nonlinear multigrid methods incorporate a full representation of the 
problem on coarse levels.  Modifications of the fine grid matrix
equation are necessary on the coarse grids
to obtain zero correction at convergence.  
These modifications are termed defect corrections.  In this paper, 
we examine the FAS multigrid method of Costiner and Ta'asan
\cite{scos95a,scos95b}. They 
presented a detailed account of the algorithm in two papers: the 
first concerns fixed potential problems and the second addresses
self-consistency. A brief summary will be given here. We will follow
the notation from their work.

\vspace{0.5cm}
\noindent
{\it FAS Eigenvalue Method}
\vspace{0.5cm}

Suppose we wish to solve an eigenvalue problem represented in real space with 
finite differences.  This leads to the matrix equation

\begin{equation}
AU = U\Lambda .
\end{equation}

\noindent
The matrix $A$ is the $N_g^h \times N_g^h$ (banded) Hamiltonian, $U$ is the
$q \times N_g^h$ matrix of the eigenvectors, and $\Lambda$ is the 
diagonal $q \times q$ matrix of eigenvalues. 

In the FAS approach, we express the coarse grid (level $i$) problem as 

\begin{equation}
F_i U_i = T_i .
\end{equation}

\noindent
Here $F_i U_i = A_i U_i - U_i \Lambda_i$ and $T_i$ is the defect 
correction.  On the finest grid $T_i = 0$. On grids $j$ coarser than $i$,

\begin{equation}
T_j = I_i^j(T_i - F_i U_i) + F_jI_i^jU_i ,
\end{equation}

\noindent
where the operator $I_i^j$ is the restriction operator. We use 
full weighting restriction throughout, which just involves a local
trapezoid-rule integration of the function values from the fine
grid.  In the above equations, if the exact numerical 
fine grid solution is 
inserted into the coarse grid equations, identities are obtained. 
This is equivalent to the zero correction at convergence 
condition.

An initial approximation is first obtained on the fine grid.
We obtain this approximation by implementing a full multigrid (FMG)
cycle \cite{jwan00}, 
beginning on the coarsest level, interpolating to the next
finer grid, performing MG cycles there, and so on until the finest
grid is reached.  In this way, a good initial approximation is 
obtained for very little numerical effort. 
Following relaxation (typically 2-5 Gauss-Seidel or 
successive over-relaxation/SOR 
steps), the fine grid approximation
is passed to the coarse level by restricting the eigenfunctions,
the potential, and the defect correction from the finer grid. 
Relaxations (and generalized Ritz projections, see below)
are performed on the current coarse grid, and the problem is
then passed again to the next coarser level.  This process is repeated
until the coarsest grid is reached. We typically utilize three grid
levels when the finest grid is a $65^3$ mesh.  Once relaxation is done
on the coarsest level, a correction step for the next finer level is
performed:

\begin{equation}
U_i^{new} = U_i^{old} + I_j^i (U_j - I_i^j U_i^{old}) .
\end{equation}

\noindent
$I_j^i$ is the interpolation operator. Linear interpolation by 
lines is used throughout except during the FMG process when 
passing to the next finer level, where cubic interpolation by 
lines is used (to obtain a better initial guess on the fine 
grid). The correction steps are continued until the finest grid
functions are corrected followed by relaxation steps there. The 
full cycle through all the levels is termed a V-cycle.
One may then repeat the MG V-cycles until a desired convergence is
obtained.

\vspace{0.5cm}
\noindent
{\it Generalized Ritz Projection, GRP}
\vspace{0.5cm}

Consider a new eigenvalue relation, in which the matrix $V$ results
from a linear combination of the current approximation $U$:

\begin{equation}
AV = V\Lambda ,
\end{equation}

\noindent
where $V=UE$ and $E$ is a $q \times q$ matrix of 
normalized vectors, the columns
of which determine the coefficients for the linear combination of 
old vectors from $U$.  Then on the fine grid we have the relation

\begin{equation}
AUE = UE\Lambda .
\end{equation}

When this problem is passed to the coarser levels, the proper FAS transfer is

\begin{equation}
AUE = UE\Lambda + TE .
\end{equation}

\noindent
If we multiply on the left by $U^T$, we obtain the following 
generalized (nonsymmetric) eigenvalue problem on the coarse grid:

\begin{equation}
U^T (AU - T) E = (U^T U) E \Lambda .
\end{equation}

\noindent
Notice that the eigenfunctions are no longer orthonormal when 
passed to the coarse grid.  Also, it is easy to show that, at 
convergence, the eigenvalues are the same on all grid levels;
in principle, there is no need to compute them on the finest grid.
We solve this $q \times q$ eigenproblem
with standard linear algebra packages.  Once solved, we obtain
new eigenvalues, eigenfunctions (linear combinations of the 
current approximation), and defect corrections (also linear 
combinations of the old defect corrections).  On the finest grid, 
where the defect correction $T$ is zero, the GRP reduces to the 
usual Ritz projection employed in the Brandt {\it et al.}
algorithm \cite{abra83}. 

\vspace{.5cm}
\noindent
{\it Backrotation, BR}
\vspace{.5cm}

A subtle aspect of the correction scheme outlined above is 
that the coarse-grid eigenfunctions must properly match 
their fine-grid counterparts for the correction step.  Therefore,
Costiner and Ta'asan \cite{scos95a} introduced a process called backrotation 
in their solver.  This operation is designed to prevent rotations
in degenerate or near-degenerate subspaces, and to prevent 
sign changes, rescalings, and permutations of the eigenvectors.
In the backrotation, the $E$ matrix is modified towards the ends listed above. 
If this step is not employed in the algorithm of 
Ref.\ \cite{scos95a}, the solver typically does not fully
converge. As a simple example, imagine that the sign of one of
the eigenfunctions changes during GRP on a coarse level.  If the
correction is then interpolated to the fine grid, the corrected
function will be severely distorted. At convergence, the $E$ matrix should 
approach the unit matrix. During processing
(prior to backrotation), it tends to have block
diagonal form, where the blocks correspond to degenerate or 
near-degenerate subspaces.  The dimensions of the blocks determine the 
eigenvalue cluster sizes.  An extensive discussion of the backrotation is
given in the original paper.     
 
\vspace{.5cm}
\noindent
{\it Relaxation}
\vspace{.5cm}

A major feature of MG methods is that relatively simple relaxation
strategies can be employed so long as they decimate 
errors with wavelengths comparable to the grid spacing on a given 
level \cite{abra77}.
Gauss-Seidel is the most common one utilized. 
We have investigated several relaxation 
strategies for smoothing on each level.  Originally, we used the 
Gauss-Seidel with shift form given in Ref.\ \cite{abra83}. 
Recently, we have extended
this relaxation method to an SOR form, and find improved 
convergence.  On the coarsest level, we have employed 
Gauss-Seidel directly, Gauss-Seidel with constraints designed to 
maintain eigenfunction orthonormality on the fine grid \cite{abra83}, and
Kaczmarz \cite{whac85} relaxation. Kaczmarz 
relaxation is guaranteed to converge, but it 
exhibits slower convergence relative to Gauss-Seidel or SOR (this is not
a significant issue on coarse levels).
It will be specified below which method 
was used for each application. 
Further details of relaxation methods will be presented
in an extensive account of our algorithm \cite{nwij03}.  

\vspace{.5cm}
\noindent
{\it Self-Consistent Problems}
\vspace{.5cm}

Some of the applications presented below concern self-consistent
solution of the Kohn-Sham equations.  In the work presented here,
we update the eigenfunctions and self-consistent potential 
sequentially.  That is, given an initial approximation to the 
effective potential, an MG V-cycle is performed to update the 
eigenfunctions.  From the new eigenfunctions, a new charge 
density is computed, from which a new effective potential 
is obtained.  The main computational step for updating the 
effective potential is solution of a Poisson problem.  This 
equation is also solved with MG V-cycles. The 
total time to solve the Poisson equation is less than that
for updating a single eigenfunction, and this operation 
scales linearly with system size. We are currently exploring
approaches to update the effective potential on coarse levels
simultaneously with the eigenfunctions \cite{scos95b}.  
This can be expected
to accelerate the self-consistent convergence rate. We note that
in our calculations so far, we have found no need for potential
or charge density mixing of old and new solutions; we believe this
is due to long-wavelength stabilization of the charge density during
the FMG preconditioning process.

\vspace{.5cm}
\noindent
{\it Pseudopotentials}
\vspace{.5cm}

For our calculations on atoms and molecules, we have incorporated
the separable dual-space Gaussian pseudopotentials developed by
Goedecker {\it et al.} \cite{sgoe96,char98}. 
These pseudopotentials have analytic forms
with only several parameters per atom, and they exhibit optimal
decay properties in both real and reciprocal space.  We have 
implemented the real-space relativistic version of these 
pseudopotentials for the present calculations.  For calculations
on the coarse grids, we simply compute the function values just as
we do on the fine level. Of course fewer grid points are necessary 
to sample the pseudopotential on coarse grids due to 
the decay properties of the projectors. It has been shown that application of 
pseudopotentials in real space is more efficient than in 
reciprocal space \cite{rkin91}.

\vspace{1.cm}
\large
\centerline{\bf Numerical Results}
\normalsize
\vspace{.5cm}

\vspace{.5cm}
\noindent
{\it Fixed Potential Problems}
\vspace{.5cm}

We first present results of the FAS algorithm on fixed potential
problems.  As a benchmark, we utilized the original algorithm of
Ref.\ \cite{abra83} and solved the same two dimensional eigenvalue 
problem addressed in that paper.  
Their kinetic energy operator is scaled by a factor of two. A second 
order finite difference approximation was assumed for the Laplacian. 
Half of their potential is

\begin{eqnarray}
v= 5y \sin(3 \pi x) .
\end{eqnarray}

The total domain size was taken as one, and four grid levels were utilized in the FAS process. 
Gauss-Seidel relaxation (with a shift parameter of zero) was employed on all four levels.
One relaxation step was performed on the coarsest level while enforcing constraints
designed to maintain eigenfunction orthonormality on the finest level. On all other levels, two 
relaxation steps were performed.
In total, 7 FAS V-cycles were implemented on the finest level. 
The computed eigenvalues and residuals are shown in Table 1.
The residual for each eigenfunction is defined as

\begin{eqnarray}
r = \sqrt{ { \sum \left | H \psi - E \psi \right |^2 } \over N_g^h} , \label{res}
\end{eqnarray}
where the sum is over all the fine grid points and $N_g^h$ is the total number of grid points.
The algorithm with Ritz projection performed on the finest level converges nicely to 
the numerically exact eigenfunction/eigenvalue pairs.  

We next compare the convergence rates of three FAS algorithms: 1) Ritz on fine grid
(Ref.\ \cite{abra83}, 2)
generalized Ritz with backrotation (GRBR) on coarse grids with no orthonormalization
on the fine level \cite{scos95a}, and 3) GRBR on coarse grids
with periodic (every 5 V-cycles) Gram-Schmidt orthonormalization on the fine grid.  These approaches
were tested on simple fixed potential problems, namely the three-dimensional harmonic oscillator
and the hydrogen atom.  Both of these physical problems have degenerate subspaces leading to 
eigenvalue clusters which must be handled in the backrotation step.  
For these problems, Gauss-Seidel relaxation was employed on all levels, except Kaczmarz
relaxation was utilized for the coarsest grid relaxation for the hydrogen atom problem
(this led to increased efficiency).  

The convergence results are presented in Figs.\ 1 and 2.  For the harmonic oscillator problem,
we solved for ten states and used a 12th-order finite difference representation for
the Laplacian. The total domain size is ten, and the fine grid is a
$65^3$ mesh; three grid levels were utilized. Atomic units are used throughout.  A shift parameter equal to the current 
eigenvalue was employed in the Gauss-Seidel relaxations (5 steps per level).  
The GRBR step was performed on the coarsest level. A total of 15 V-cycles were
conducted. The compute times per V-cycle for the three algorithms listed above were 50.9,
34.6, and 36.6 sec., respectively, on an 800 MHz machine.  
The $q$ values are small enough that the orthogonalization
and Ritz projections do not yet dominate the compute time. 
For the hydrogen atom case, we generated the fixed potential by numerical solution 
of the Poisson equation for a fixed central unit charge.  Again, 12th-order finite differences 
were employed, and we solved for 14 states on a domain with a total side length  of 28. The fine grid is a $65^3$ mesh. For this 
case, it was found that a shift parameter equal to the potential was more efficient. A total 
of 20 V-cycles were conducted.  For both cases, the fastest (lowest eigenvalue) and slowest 
(highest eigenvalue) converging cases are shown.  The GRBR (without orthonormalization) operation was conducted
on the middle grid level, while for the third algorithm with periodic Gram-Schmidt operations 
on the fine grid, the GRBR step was implemented on the coarsest level.  The compute times per 
V-cycle were 81.5, 59.7, and
52.8 sec.\ for the three algorithms. 

For both physical problems, performing the orthonormalization and Ritz projection on the 
fine level leads to the most efficient convergence.  The harmonic oscillator potential is 
smooth, and the GRBR algorithm exhibits good convergence behavior. However, the convergence rate is slightly
slower than when orthonormalization and Ritz projection are performed on the finest level.
The GRBR convergence rate is the same whether or not orthonormalization is periodically
done on the finest level, indicating good separation can be obtained without processes on 
the fine level. For the hydrogen atom case, the overall convergence rate is reduced 
relative to the harmonic oscillator, presumably due to the singular nature of the potential.
Also, periodic Gram-Schmidt operations on the finest level increase the convergence rate 
slightly.  

As part of the backrotation step, the degenerate subspaces (eigenvalue clusters) must be identified so as
to prevent rotations within those clusters.  Both the harmonic oscillator and the hydrogen
atom problems possess clear eigenvalue structure
which can be directly incorporated or determined during the solution process.  
This ensures healthy convergence of the GRBR 
algorithm, behavior which was also observed in the original work of 
Costiner and Ta'asan \cite{scos95a}.  We will see below that, for 
large self-consistent molecular 
cases without such well-defined 
eigenvalue cluster structure, the algorithm may stall due to mixing during the backrotation step.  

\vspace{.5cm}
\noindent
{\it Self-Consistent Pseudopotential Calculations}
\vspace{.5cm}

        Computational results for self-consistent 
pseudopotential Schr\"odinger-Poisson 
eigenvalue problems are presented. Just as for the fixed potential problems 
presented above, the 
algorithms used were Ritz projection on the fine level coupled with Gram-Schmidt 
orthogonalisation and the GRBR algorithm without and with periodic 
fine grid Gram-Schmidt operations. In both GRBR algorithms, the eigenfunctions
were normalized on the fine level to ensure charge conservation.
Three examples,  Ne, CO, and the benzene dithiol molecule were used in the study. All three 
cases are three dimensional and were treated with 12th order finite
difference representations. Three grid 
levels were utilized comprising $17^{3},33^{3}$ and $65^{3}$ total 
points . The first 
example, the Ne atom, possesses 4 occupied states. 
Similarly 5 and 21 eigenvectors were 
required for the CO and benzene dithiol molecules, respectively. The Ne and CO
examples possess well-defined eigenvalue cluster structure (triply degenerate p
states for Ne and a doubly degenerate $\pi$ bonding orbital for CO). 
Convergence results are presented in Figs.\ 3-5.

        Choosing the optimal parameters for the relaxation scheme was 
important. The shift parameter for Gauss-Seidel relaxation was
taken as $\lambda + v$, where $\lambda$ is the eigenvalue and 
$v$ is symbolically the effective potential 
(nonlocal in the case of the pseudopotential). Gauss-Seidel relaxation 
($\omega = 1$, where $\omega$ is the overrelaxation parameter) 
was used on all levels except on the finest grid where SOR relaxation was 
employed ($\omega = 1.7$). These near-optimal 
relaxation parameters were determined
by numerical experiments. In the final V-cycle involving 
the three grid levels, three relaxation steps were done on each level. 

All three algorithms converged for the Ne and CO cases.  The
fine grid Ritz plus Gram-Schmidt and coarse-grid GRBR with periodic
fine level Gram-Schmidt algorithms both exhibited excellent 
convergence rates.  Implementing GRBR with no fine grid separation (orthogonalization)
led to slower convergence; the total energy 
convergence slows after a few self-consistency iterations.  
In the GRP, the fine level separation of 
wavefunctions comes as a result of GRBR done on the middle level. But complete 
separation may not always be achieved on the fine level (Table 2). Performing
only a few fine grid Gram-Schmidt operations during the entire solution process
restores the convergence rate and leads to acceptable fine grid eigenfunction 
orthogonality at the end.  Clearly the form of the potential (self-consistent
pseudopotential in this case) and resulting eigenvalue structure have 
impacts on the convergence behavior
of the GRBR algorithm.  

        In the case of benzene dithiol, the fine grid separation algorithm
converged nicely as for the smaller molecules. 
However, neither variant of the GRBR algorithm converged properly; the solver
stalled with only modest energy convergence. 
To probe for the reason for this lack of convergence, we first converged the 
system using fine-grid separation for 15 V-cycles, and then used the resulting potential in 
a fixed potential calculation.  We observed that the $E$ matrix determined from
GRBR gradually begins to destablize rather than to converge to the unit matrix
as expected.  This suggests again that the form of the potential and the resulting
eigenvalue cluster structure affect the convergence of the GRBR algorithms.
Benzene dithiol is a relatively  
large molecule with 14 atoms (C,O,S and H) and 21 valence 
eigenstates. We believe that
the difficulties arise from the ambiguous cluster structure of the eigenvalues
which leads to mixing during the backrotation operation. Similar
calculations were performed on the benzene molecule, which possesses clear 
symmetry and cluster structure, and the algorithm converged.
Since determination of 
the cluster structure is crucial for convergence, and this must be performed
automatically in the algorithm in order to treat general systems, this issue
must be addressed for the GRBR approach to provide a generally convergent scheme.
Work is in progress investigating these difficulties.  One possible solution
is suggested below.  

\vspace{1.cm}
\large
\centerline{\bf Discussion}
\normalsize
\vspace{.5cm}

The objective of this paper has been to provide a test of the 
nonlinear FAS multigrid eigenvalue method of Refs.\ \cite{scos95a,scos95b}
for solving the Kohn-Sham equations.  This method is promising 
because it removes the expensive orthogonalization and Ritz 
projection operations to coarse levels (with a corresponding 8-fold
reduction in cost per level in three dimensions).  The necessary fine
grid work only involves relaxation, normalization of the eigenfunctions,
and perhaps orthogonalization within degenerate clusters.
The model 
problems treated in Refs.\ \cite{scos95a,scos95b} possess relatively
smooth potentials and well-defined eigenvalue cluster structure.  Similar
to the results of Costiner and Ta'asan, we find good convergence of the 
GRBR approach for fixed potential 
and self-consistent eigenvalue problems with well-defined
eigenvalue clusters. However, we found that for a larger molecular case with
complicated eigenvalue structure, the GRBR approach did not converge properly.
We linked these difficulties to the backrotation step which appears to be 
highly sensitive to the determination of the clusters. Since determination of
the clusters must be done numerically during the course of the solution 
process, this issue must be addressed to develop a generally convergent solver
for large systems.

We have recently begun investigating one possible approach to deal with
these difficulties.  Notice that the $E$ matrix of Eqs.\ 6 and 7 is  
formally the same on all levels at convergence (just as are the eigenvalues).
In the algorithm of Ref.\ \cite{scos95a}, the backrotation involves a 
modification of the $E$ matrix designed to prevent rotations in degenerate
or near-degenerate subspaces, sign changes, rescalings, and permutations
of eigenvectors.  Utilizing the fact that the $E$ matrix is the same on 
the coarse and fine levels (at convergence), we can circumvent the backrotation
by first applying the $E$ matrix to the current fine-grid approximation to 
the eigenfunctions prior to the correction step.  This approach is in a sense
a hybrid of the fine and coarse grid Ritz projections; we use the coarse grid
to generate the new eigenvalues and the $E$ matrix, but we use that $E$
matrix to alter the fine-grid occupied subspace.  Thus the expensive step of
constructing the Ritz matrix has been moved to the coarser level.  The use of the
$E$ matrix to update the fine-grid function is a relatively cheap operation. Formally,
it does scale as $q^2N_g^h$, but the $E$ matrix is of block diagonal form, with 
the blocks of dimension of the corresponding degenerate cluster.  These clusters
are generally very small.  Therefore realistically the update operation scales
as $qN_g^h$ if the eigenfunctions span the whole domain.  
Some discussion along these lines was already given in Ref.\ \cite{abra83}.
We are currently exploring
the use of this idea in our nonlinear FAS multigrid solver.  
In preliminary results, 
we have found it to converge properly for all of the physical 
problems examined in this paper. 

\vspace{1.cm}
\large
\centerline{\bf Acknowledgments}
\normalsize
\vspace{.5cm}

We gratefully acknowledge the support of the National Science Foundation
(CHE-0112322) for this research.  We also thank Achi Brandt and 
Shlomo Ta'asan for many helpful discussions.

\newpage
%%%%%%%%%%%%%%%%%%%%%%%%%%%%%%%%%%%%%%%%%%%%%%%%%%%%%%%
% REFERENCES
%%%%%%%%%%%%%%%%%%%%%%%%%%%%%%%%%%%%%%%%%%%%%%%%%%%%%%%

\newpage

\begin{table}[h]
\vspace{2mm}
\centerline{
\begin{tabular}{|l||l|}
\hline
$Table\; 1\quad$ Ritz result &  \quad Ref.\ \cite{abra83} \\
\hline
$No. \quad \quad \quad 2\lambda \quad \quad \quad r$ & $\quad \lambda \quad\quad $\\
\hline \hline
$1 \quad 18.71847149 \quad 3.3E-11$ & $18.71847149 $\\
$2 \quad 48.18927363 \quad 4.1E-09$ & $48.18927363 $\\
$3 \quad 51.56004355 \quad 5.0E-12$ & $51.56004355 $\\
$4 \quad 81.07201016 \quad 2.8E-11$ & $81.07201016 $\\
$5 \quad 97.00117915 \quad 9.6E-09$ & $97.00117915 $\\
$6 \quad 99.57484220 \quad 1.2E-08$ & $99.57484220 $\\
$7 \quad 129.1084354 \quad 1.7E-06$ & $129.1084354 $\\
$8 \quad 129.8996943 \quad 2.5E-07$ & $129.8996943 $\\
\hline
\end{tabular}
}
\caption{Comparison of eigenvalues to results of Ref.\ \cite{abra83}.}
\vspace{2mm}
\end{table}

\newpage
\begin{table}[htb]
\begin{center}
\begin{tabular}{|c||c|c|c||c|c|c|}

\hline
  & CO &  &  & Ne &   &  \\
\hline

\hline
 Wfs & GS & GRBR & GRBR+GS & GS & GRBR & GRBR+GS\\
\hline

\hline
(wf0*wf0) & 1.0e+00   & 1.0e+00   & 1.0e+00   & 1.0e+00   & 1.0e+00   & 
1.0e+00  \\
(wf0*wf2) & -1.4e-17  & -7.3e-07  & -5.6e-08  & -3.4e-17  & -1.2e-07  & 
9.3e-16  \\
(wf0*wf3) & -6.2e-18  & 3.9e-07   & -1.9e-08  & 5.0e-18   & -4.6e-07  & 
2.2e-12  \\
(wf1*wf2) & -1.7e-17  & 1.7e-06   & -6.9e-09  & 1.3e-16   & 2.0e-05   & 
-6.0e-11 \\
(wf1*wf3) & 3.0e-17   & -5.0e-07  & 9.3e-08   & -1.8e-18  & -2.7e-04  & 
7.8e-10  \\
(wf2*wf3) & 9.5e-17   & -5.9e-04  & 1.2e-06   & 7.4e-18   & -2.8e-05  & 
6.9e-11  \\
(wf0*wf4) & -1.9e-18  & 2.9e-08   & 2.9e-09   & -         & -         & 
-        \\
(wf1*wf4) & -2.4e-18  & 2.1e-08   & 3.4e-08   & -         & -         & 
-        \\
(wf2*wf4) & -1.3e-19  & -1.4e-07  & 4.0e-07   & -         & -         & 
-        \\
(wf3*wf4) & -5.1e-19  & 4.4e-08   & 3.8e-09   & -         & -         & 
-        \\

\hline

\hline
\end{tabular}
\caption{A sample of dot products of wavefunctions of CO and Ne are 
shown. Column GS includes products when the Gram-Schmidt 
orthogonalization was performed and it serves only as a reference. The column 
GRBR includes the products as a result of GRBR only and the wavefunctions 
are not fully separated. The column GRBR+GS includes the products 
when Gram-Schmidt orthogonalization was perfomed at a regular interval 
(5,10,15 V-cycles only). The separation of wavefunctions is improved. The 
total number of V-cycles perfomed was 20.}
\end{center}
\end{table}

\newpage

\begin{figure}[h]
%\centerline{ \epsfxsize=7cm \epsfysize=5cm \epsfbox{ HO.eps } }
\centerline{ \epsfxsize=12cm \epsfysize=10cm \epsfbox{ 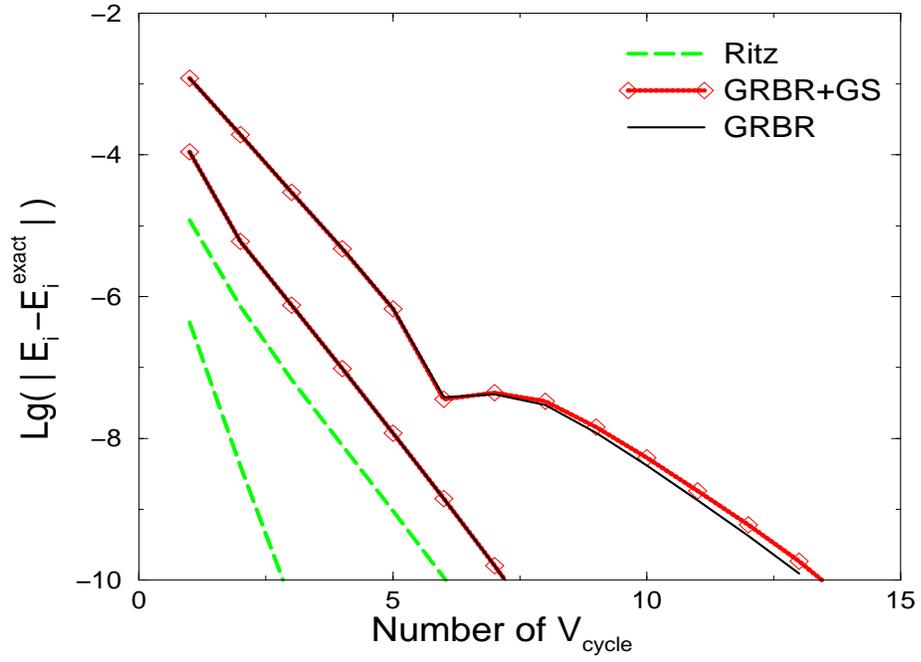 } }
\vspace {3 mm}
\caption{Convergence rates for different methods in a fixed harmonic oscillator potential. 
Long dashed lines are the results of the Ritz method on the fine grid, 
the solid lines are results of the GRBR method on the coarsest level, and the dotted lines with diamonds are results of 
the GRBR method on the coarsest level with additional 
periodic Gram-Schmidt orthogonalization on the fine grid.}
\label{hofig}
\end{figure}

\begin{figure}[h]
%\centerline{ \epsfxsize=7cm \epsfysize=5cm \epsfbox{ Hatom2.eps } }
\centerline{ \epsfxsize=12.cm \epsfysize=10.cm \epsfbox{ 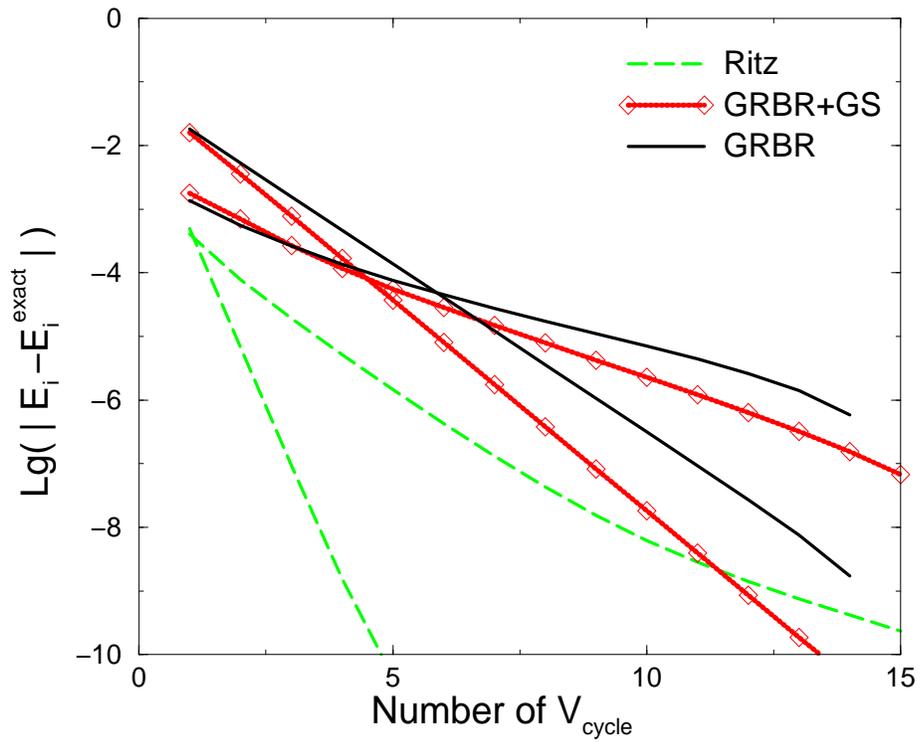 } }
\vspace {3 mm}
\caption{Convergence rates for different methods in a fixed hydrogen atom potential.  
Long dashed lines are the results of the Ritz method on the fine grid, the solid lines are the results of the GRBR method 
on the $33^3$ grid, and the dotted lines with diamonds are results 
of the GRBR method on the $17^3$ grid with additional periodic 
Gram-Schmidt orthogonalization on the fine grid.} 
\label{hatomfig}
\end{figure}

\newpage

\begin{figure}
\begin{center}
\begin{minipage}{10.0cm}
%\resizebox{8.0cm}{5cm}{\includegraphics{ne.eps}}
\centerline{\resizebox{12.0cm}{10.0cm}{\includegraphics{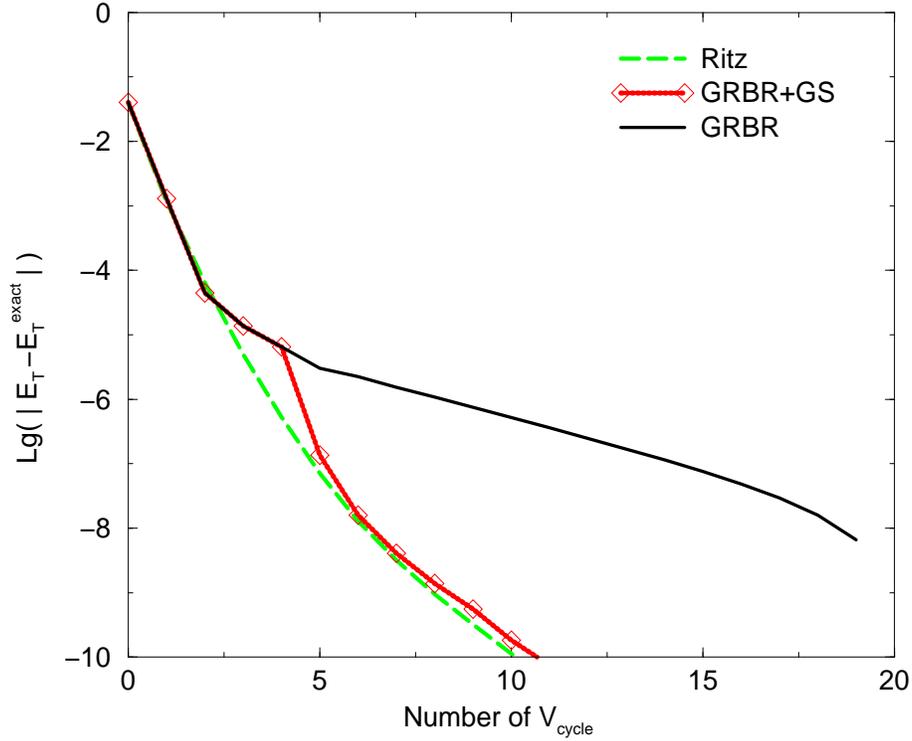}} }
\end{minipage}
\caption{Convergence rate for Ne. The logarithm (base 10) of
the difference between the current and fully converged
total energies is plotted against 
the number of V-cycles (self-consistency steps). 
Ritz and GRBR stand for 
fine-grid Ritz projection and 
coarse-grid generalised Ritz projection, respectively. 
GRBR-GS is  for GRBR with 
fine-grid Gram-Schmidt orthogonalization performed at 3 V-cycles 
(5,10,15). The fine grid spacing used was h=0.178437.}
\end{center}
\end{figure}

\newpage

\begin{figure}
\begin{center}
\begin{minipage}{10.0cm}
%\resizebox{8.0cm}{5cm}{\includegraphics{co.eps}}
\centerline{\resizebox{12.0cm}{10.0cm}{\includegraphics{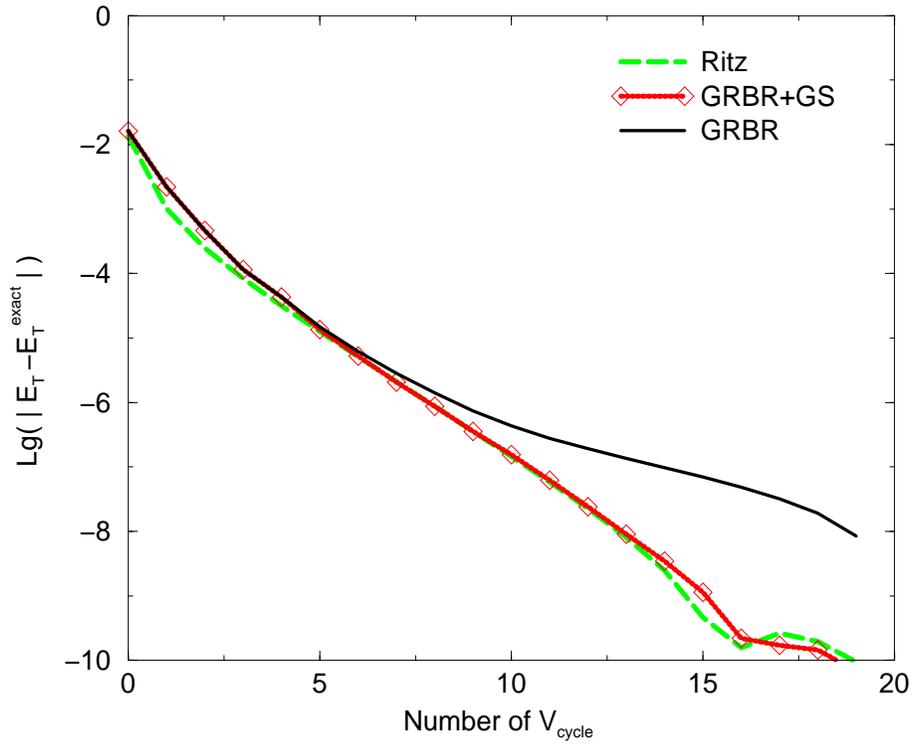}} }
\end{minipage}
\caption{Convergence rate for CO. The logarithm (base 10) of the difference
between the current and fully converged total energies is plotted against 
the number of V-cycles (self-consistency iterations). 
Ritz and GRBR stand for fine grid Ritz projection and
coarse-grid 
generalized Ritz projection with backrotation, respectively. 
GRBR-GS is for coarse-grid GRBR 
with Gram-Schmidt orthogonalization performed 
on the fine grid at 3 V-cycles 
(5,10,15). The fine grid spacing used was h=0.178437.}
\end{center}
\end{figure}

\newpage

\begin{figure}
\begin{center}
\begin{minipage}{10.0cm}
%\resizebox{8.0cm}{5cm}{\includegraphics{bd1.eps}}
\centerline{\resizebox{12.0cm}{10.0cm}{\includegraphics{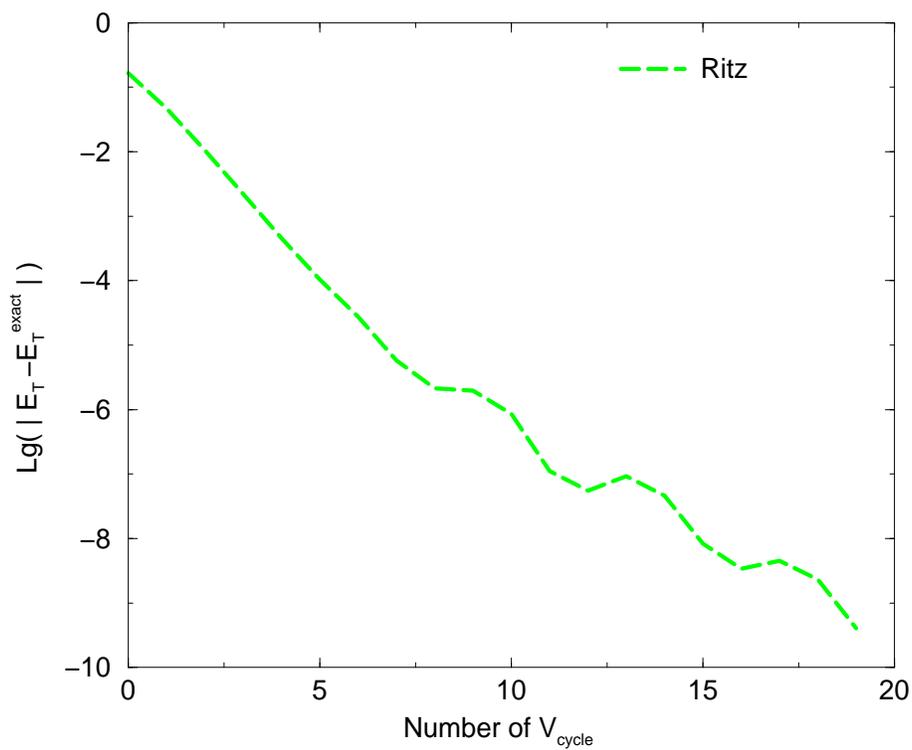}} }
\end{minipage}
\caption{Convergence rate for benzene dithiol. The logarithm (base 10)
of the difference between the current total energy and the 
fully converged value is plotted 
against the number of V-cycles (self-consistency iterations). 
Only the fine grid 
Ritz projection case is shown. The 
fine grid spacing used was $h$=0.3.}
\end{center}
\end{figure}

\end{document}